\documentclass[aps,twocolumn]{revtex4}
\usepackage{amsfonts}
\usepackage{amssymb}
\usepackage{amsmath}
\usepackage{graphicx}
\usepackage{epsfig,amsmath}
\usepackage[bookmarksnumbered,linktocpage,pdfstartview=FitH]{hyperref}

\begin{document}

\title{Thermodynamics of random walking vortex loops in counterflowing superfluids}
\author{Sergey K. Nemirovskii\thanks{%
email address: nemir@itp.nsc.ru}}
\affiliation{Institute of Thermophysics, Lavrentyev ave, 1, 630090, Novosibirsk, Russia\\
and Novosibirsk State University, Novosibirsk}
\date{\today }

\begin{abstract}
Based on the theory of the thermodynamic equilibrium in a system of quantum
vortices in superfluids in the presence of a counterflow, the influence of a vortex tangle on various thermodynamic phenomena in quantum liquids is studied. Using the early calculated partition function we study some of the properties of He II related to counterflow, such as the distribution of vortex loops in their length, the suppression of the superfluid density $\rho _{s}$ and the shift $T_{\lambda}$. Good agreement with the early obtained results is a fairly strong argument in favor of the point of view that the gas of string-like topological excitations can indeed be considered as a additional kind of quasi-particles having the inner structure at high temperatures, especially near the phase transition. The application of the developed formalism to the theory of quantum turbulence is briefly discussed.
\end{abstract}

\maketitle

\nopagebreak

\section{Introduction and scientific background}

Long ago, in pioneering papers by Onsager \cite{Onsager1949b} and by Feynman
\cite{Feynman1955} the idea that the quantum vortices might be
responsible for the superfluid phase transition had been suggested. Since
then, many authors developed this idea\ by studying the gas of string like
topological excitations in various media. Of course, the main achievements
were in the two-dimensional case, where the
Berezinskii--Kosterlitz--Thouless transition was predicted and thoroughly
investigated \cite{Kosterlitz2017} As for 3D case, the situation here is far
from a more less developed theory. Nevertheless, there are quite a few works
that use various approximations \cite{Kleinert1991},\cite{Kleinert1990a},%
\cite{Vilenkin1994}, \cite{Copeland1991}, \cite{Williams1999},\cite{Lund1990}%
,\cite{Chorin1994}). In these papers, the effect of an ensemble of vortex
filaments on the phase transition, on the behavior of the superfluid
component, on the behavior of the heat capacity, and other issues were
discussed. It can be stated \cite{Williams1999},\cite{Copeland1991},\cite%
{Vilenkin1994} that the theory of string-like structures can provide a
physical understanding of phenomena in superfluid liquids, superconductors,
models of cosmic strings, crystalline bodies (melting), pulsars (glitches).\newline
In the letter we apply this ideas to the counterflowing superfluids. There
are many works in which various properties of superfluidity in the presence
of counterflow are investigated. It is of interest to consider these
phenomena from the standpoint of an ensemble of quantum vortices.\newline
In the paper we will focus on the effects related to counterflow. In
particular we will study the distribution of vortex loops in the space of
their lengths, the vortex line density (VLD) $\mathcal{L}$ total length per
unit volume, the Lamb impulse and the superfluid flow induced by vortices.
We also investigate suppression of superfluid density $\rho _{s}$ and shift $%
T_{\lambda }$\ (which we associate with Hagedorn temperature $T_{H}$\ ) due
to the counterflow velocity. Additionally, we discuss possible
applications of thermodynamically equilibrium vortices to the theory of
quantum turbulence (QT).

\section{Gibbs distribution and partition function}

In paper by the author \cite{Nemirovskii2016Vns} it was shown that a vortex
filaments in superfluids under the action of a random Langevin force in the
presence of a counterflow with a relative velocity $\mathbf{v}_{ns}$ come
into a state of thermodynamic equilibrium with the Gibbs distribution
\begin{equation}
\mathcal{P}(\{\mathbf{s}(\xi )\},t)=\mathcal{N}\exp (-\frac{H\{\mathbf{s\}}}{%
k_{B}T}).  \label{Gibbs}
\end{equation}%
Here $\mathbf{s(}\xi \mathbf{,}t\mathbf{)}$ are the radius vectors of the
vortex line elements, $\xi $ is a label parameter, in this case it coincides
with the arc length, and, accordingly, runs from $0$ to the length of the
loop $l$; Quantity $\mathcal{N}$ is a normalization factor. The Hamiltonian
in the presence of relative velocity has the form%
\begin{equation}
H\{\mathbf{s\}=}E\{\mathbf{s\}-P\cdot \mathbf{v}}_{ns}\mathbf{.}
\label{H(s)}
\end{equation}%
The energy $E\{\mathbf{s\}}$ and the Lamb impulse $\mathbf{P}$ are defined
as (see, for example, \cite{Nemirovskii2013})
\begin{equation}
E(\mathbf{s})=\frac{{\rho }_{s}{\kappa }^{2}}{8\pi }\int_{0}^{l}\int_{0}^{l}%
\frac{\mathbf{s}^{\prime }(\xi )\mathbf{s}^{\prime }(\xi ^{\prime })}{|%
\mathbf{s}(\xi )-\mathbf{s}(\xi ^{\prime })|}d\xi d\xi ^{\prime },\ \ \
\label{E and P}
\end{equation}%
\begin{equation}
\ \mathbf{P}\ \mathbf{=}\frac{\rho _{s}\kappa }{2}\int_{0}^{l}\mathbf{s}(\xi
)\times \mathbf{s}^{\prime }(\xi )\ d{\xi ,}  \label{P}
\end{equation}%
where $\kappa $ is the quantum of circulation, and ${\rho }_{s}$ is
superfluid density. Relations (\ref{Gibbs}) - (\ref{E and P}) should be used
to calculate the partition function (PF), and, accordingly, to determine the
various properties of the vortex tangle (VT). In the ordinary statistical
mechanics when $\mathbf{P}$ is the true momentum of a particle (or
quasiparticle), Eqs. (\ref{Gibbs}-\ref{H(s)}) are obvious and follows from
the Galilean transformation. Unlike this, the situation for vortex loops is
more complicated, since the Lamb impulse is not the \textquotedblright
real\textquotedblright\ momentum, and Eqs. (\ref{Gibbs}) - (\ref{P}) need
the verification, which had been accomplished in \cite{Nemirovskii2016Vns}%
.In the next work by the author devoted to this topic \cite{Nemirovskii2022}
the one loop PF $Z_{1}$, corresponding the stated problem (\ref{Gibbs}) - (%
\ref{P}) was calculated. In this calculation the vortex lines were supposed
to be closed loops having a random walking structure on the cubic lattice
with edge \ equal to $a$ and obeying the Wiener distribution. The starting
expression for $Z_{1}$ reads:
\begin{widetext}
\begin{gather}
Z_{1}=\int d\mathbf{r}\sum_{l}\left( \frac{a}{l}\right) \int_{\mathbf{s}(0)=%
\mathbf{r}}^{\mathbf{s}(l)=\mathbf{r}}D\mathbf{s}(\xi \}\exp \left[ -\frac{3%
}{2a}\oint (\mathbf{s}^{\prime }(\xi ))^{2}d\xi +\ln 5\frac{l}{a}\right]
\notag \\
\times \exp \left[ -\beta \frac{{\rho }_{s}{\kappa }^{2}}{8\pi }%
\int_{0}^{l}\int_{0}^{l}\frac{\mathbf{s}^{\prime }(\xi _{1})\cdot \mathbf{s}%
^{\prime }(\xi _{2})}{\left\vert \mathbf{s}(\xi _{1})-\mathbf{s}(\xi
_{2})\right\vert }d\xi _{1}d\xi _{2}+\beta \mathbf{v}_{ns}\frac{\rho
_{s}\kappa }{2}\int_{0}^{l}\mathbf{s}(\xi )\times \mathbf{s}^{\prime }(\xi
)\ d{\xi }\right] .  \label{Z1}
\end{gather}%

\end{widetext}The first line is the contribution into PF due to
configurations of vortex loops. The path integral take into account all
curves starting and ending in the same point $\mathbf{r}$. Term $(l/a)\ln 5$
appears due to fact that $l/a$ links, placed on the cubic lattice, have $%
(2D-1)$ options for which the string can move at each step, so the number of
line states $N(l,a)$ increases with its length as $N(l,a)=(2D-1)^{l/a}$ ( $D$
is the dimension of space). The quantity $\ln 5$\ can be considered as the
entropy (normalized by the Boltzman constant $k_{B}$) per link. It is
interesting that for polymer chains the entropy per leg, obtained in
numerical simulations (see \cite{Chorin1994}) is about $1.6$, which is very
close to $\ln 5$. The quantity $a$ is the length parameter of model. For the
cubic lattice model, it is the cube edge; for polymer chains it is an
elementary step. In the case of quantum vortices, quantity $a$ is the
coherence length (see for details, e.g., \cite{Kleinert1991},\cite%
{Kleinert1990a}, \cite{Copeland1991}). Further the integration over $\mathbf{%
r}$ will be omitted, since the spacial homogeneity is assumed and,
accordingly, $Z_{1}$ is considered as it is taken per unit volume.\newline
In work \cite{Nemirovskii2022} \ the calculation of the PF was implemented on the
base of the local induction approximation (LIA) for the energy $E(\mathbf{s}%
) $. In LIA, the energy of the vortex loop is $E_{loc}=\varepsilon _{V}l$,
where $\varepsilon _{V}$ is the energy of a vortex line per unit length
(tension, see, e.g. \cite{Donnelly1991}). The main achievement of the LIA is
that the path integral in \ref{Z1} becomes Gaussian and can be calculated
exactly. Realizing that a complete description should be carried out on the
basis of a complete expression for the energy $E$\ (\ref{E and P}), we would
like note, however, that the LIA is capable of describing many important
features of the vortex ensemble. It can be motivated by that the main part
of the dynamics is described precisely by the LIA (See, e.g. review article
\cite{Nemirovskii2013}). And, more importantly, since the main goal of our
investigations is to study the effects related to the counterflow, the LIA
is convenient option for this purpose.\newline
The further calculations are close to the problem of the motion of a charged
particle in a constant magnetic field \cite{Kleinert1991},\cite%
{Kleinert1990a}, although there are some distinctions and a bit different
methods have been used. Result of according calculations (when $v_{ns}$ is
directed along $z$) is that the following expression for the one loop PF
(see \cite{Nemirovskii2022})

\begin{equation}
Z_{1}=\int \frac{dl}{l}\left( \frac{3}{2\pi la}\right) ^{3/2}\left[ \frac{%
\sin (l\kappa \beta a\rho _{s}v_{ns}/6)}{l\kappa \beta a\rho _{s}v_{ns}/6}%
\right] ^{-1}\exp \left[ -\beta \tilde{\varepsilon}l\right]  \label{Z1fin}
\end{equation}

Here we have introduced quantity $\tilde{\varepsilon}=(\varepsilon _{V}-%
\frac{\ln 5}{\beta a})$. The striking difference from the Kleinert's result
is that we have an ordinary sine\ instead of a hyperbolic one. Formally this
difference arises from the imaginary unit in the (Euclidean) path integral
in the expression for the motion of a particle in a magnetic field.
Physically the sine is responsible for instability of the flow with respect
to formation of large loops. Indeed, the combination $E\{s\}$ $-$ $P\cdot
\mathbf{v}_{ns}$ is just an analog of the Landau criterion (but not for
quasiparticles, but for large vortex loops).\newline
Developing the factor in square brackets into a power series in the variable
$v_{ns}$ (up to second term, the first one is obviously equal to unity), we
split the one loop PF $Z_{1}$ (\ref{Z1fin}) in two parts - the velocity-free
PF $Z_{1free}$ and the velocity dependent part $Z_{1Vns}$. \newline
The quantity $Z_{1}$ ts the PF for a single closed string of arbitrary
length. Consistently assuming that in LIA vortex loops \ do not interact, \
we calculate the grand canonical PF $Z$ for an ensemble of many loops using
the following simple formula

\begin{equation}
Z=\sum\limits_{N=0}^{\infty }\frac{Z_{1}^{N}}{N!}=\exp \left[ Z_{1}\right] .
\label{ZviaZ1}
\end{equation}%
The quantities $Z$, $Z_{1}$, $Z_{1free}$ and $Z_{1Vns}$ \ are the key ones
for calculating various properties of superfluids due to the presence of a
VT. Let's discuss a series of issues.

\section{Various applications}

\subparagraph{Two populations}

Loosely speaking, there are two ingredients that affect PF. The first one is
associated with thermal activation. The dynamics of such vortices, which we
will call thermodynamic vortices, is determined mainly by the exponential
(Gibbs) factor in (\ref{Z1fin}), and their sizes are concentrated in the
vicinity of the correlation length $a.$ The role of the velocity enclosed in
square brackets in (\ref{Z1fin}), is reduced to small corrections to the
structure of the VT and to the thermodynamic properties of the system.%
\newline
There exists one more population of vortex loops, which we call hydrodynamic
vortices. Their appearance is quite similar to nucleation of free vortices
from bound pairs in the 2D case \cite{Ambegaokar1980}. From the form of the
Hamiltonian $H\{\mathbf{s\}}$ and from the saddle-point approximation, it
follows that this population is concentrated near the scales where the
energy $E\{\mathbf{\ s\}}$ competes with the Lamb impulse $\mathbf{P}$
multiplied by $\mathbf{v}_{ns}$ (see Eq. (\ref{Gibbs})-(\ref{E and P})). It
is seen that the most preferred scale is of the order $\kappa /v_{ns}$. This
result is close to what takes place in the theory of quantum turbulence. A
full consideration of such a situation seems very complicated and is beyond
the scope of this article. In this article, we will focus only on
thermodynamic vortices and their influence on the properties of superfluids.
Questions related to hydrodynamic vortices, as well as other questions
relating to the relationship of equilibrium quantum vortices with quantum
turbulence (QT), are of great interest and will be investigated in the
future.

\subparagraph{The Hagedorn temperature}

It can be seen that for $\tilde{\varepsilon}<0$. the PF $Z_{1}$ (\ref{Z1fin}%
) diverges. The corresponding limiting temperature $T_{H}$, when $\tilde{%
\varepsilon}(T_{H})=0$ is called the Hagedorn temperature \cite{Kleinert1991}%
,\cite{Kleinert1990a}, \cite{Copeland1991}, \cite{Antunes1998} and is
defined as%
\begin{equation}
T_{H}=\frac{\varepsilon _{V}a}{k_{B}\ln 5}.  \label{TH}
\end{equation}%
In paper \cite{Antunes1998} it was shown that the $T_{H}$\ is practically
equal to the phase transition temperature ($T_{\lambda }$\ in the case of He
II). A small difference arises due to the finiteness of the system and
disappears at large volumes. The physical meaning of $T_{H}$ is that the
configurational entropy (multiplied by temperature) is comparable to the
energy of a vortex filament link. Therefore, the energy entering the system
is spent on the increasing the length of the lines, and the phase transition
is accompanied by the appearance of very long filaments.

\subparagraph{Vortex line density. A velocity-free state $\mathbf{v}_{ns}=0$.%
}

A velocity-free state \ describes thermodynamically equilibrium vortices in
helium at rest. Previously, this problem was studied under various
assumptions. The most popular were studies in which vortex filaments were
considered as ideal rings (see, for example, \cite{Williams1999},\cite%
{Lund1990},\cite{Chorin1994}). Another approach used in our article, when
vortex filaments were either flexible polymers with an effective link, or
lattice models were used \cite{Kleinert1991},\cite{Kleinert1990a}, \cite%
{Copeland1991}, \cite{Burakovsky2000}.\ One more direction, was the study a
statistics of topological defects of the underlying fields (see, e.g., \cite%
{Antunes1998}). In these papers the authors predicted that VLD $\mathcal{L}$
at the Hagedorn temperature $T_{H}$ has a value of the order of the inverse
squared coherence length, that is, $\mathcal{L\ }\symbol{126}\ (1/a)^{2}$.
Then $\mathcal{L}$ is exponentially suppressed below $T_{H}$. All these
facts are immediately follows from Eq. (\ref{Z1fin}) for PF (if we put $%
\mathbf{v}_{ns}=0$). Indeed, the integrand is the production of the power
and exponential ingredients. Therefore, at the Hagedorn temperature $T_{H}$,
when $\tilde{\varepsilon}=0$ the loops distribution becomes a pure power
function $n(l)=\left( \frac{1}{l}\right) ^{5/2}\left( \frac{3}{2\pi a}%
\right) ^{3/2}$, which indicates the random walking nature of thevortex line
structure. For the temperatures $T<T_{H}$, when $\tilde{\varepsilon}>0$, the
Gibbs factor prevails and long loops are exponentially suppressed.\newline
\begin{figure}[tbp]
\includegraphics[width=7cm]{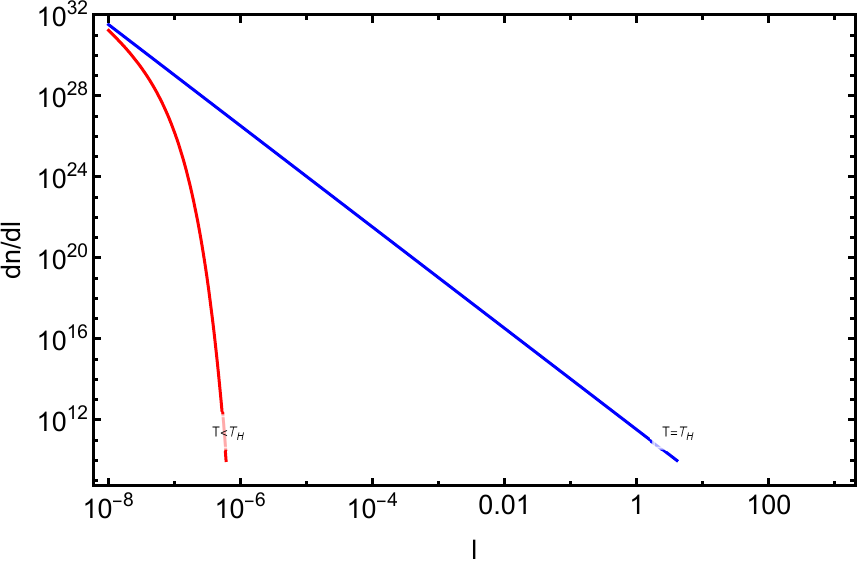}
\caption{(Color online) The loop length distribution for the Hagedorn
temperature $T_{H}$ and for $T<T_{H}$, see text, In this figure, we do not
take into account the divergence of the correlation length near the lambda
point.}
\label{dnld}
\end{figure}
Thus, we can state, that our results in the velocity-free state are in good
agreement with those previously obtained.

\subparagraph{Vortex line density due to counterflow $\mathcal{L}(\mathbf{v}%
_{ns})$}

Let's come to effects related to presence of the counterflow, $\mathbf{v}%
_{ns}\neq 0$. We begin with the correction for the VLD $\mathcal{L}(\mathbf{v%
}_{ns})$. It can be evaluated from the velocity dependent part $Z_{1Vns}$ of
PF $Z_{1}$ (\ref{Z1fin}). The most important result is that the VLD is
proportional $v_{ns}$ squared, i.e. $\ \mathcal{L}_{therm}(v_{ns})\mathcal{%
=\gamma }_{therm}^{2}v_{ns}^{2}$. \ It is noteworthy that this result is
also valid while using the sine (in Eq. \ref{Z1fin}) without a power series
developing. Quantity $\mathcal{\gamma }_{therm}$, calculated from $Z_{1Vns}$%
, as a function of the temperature, is shown in the inset of Fig. 2.

It is remarkable fact that $\mathcal{\gamma }_{therm}$ is close to the $%
\mathcal{\gamma }$, which takes place in the theory of quantum turbulence, $%
\mathcal{L}_{turb}\mathcal{=\gamma }_{turb}^{2}v_{ns}^{2}$ (see, e.g., \cite%
{Kondaurova2014}). However, in other aspects, the set of thermodynamic
vortex loops related to $Z_{1Vns}$ has properties different from the loops
observed in the theory of quantum turbulence. In particular, the average
loop length $\left\langle l\right\rangle $ is of the order of the coherence
length $a$, in contrast to the QT state, where $\left\langle l\right\rangle
\sim 1/\sqrt{\mathcal{L}}$.This is understandable, since the main source of
length is the thermal effect, and $k_{B}T/\varepsilon _{V}\backsim a$. \
Furthermore, the interline space $1/L_{therm}(v_{ns})$\ \ far exceeds the
radius of curvature and sizes of loops (which are of the order of $a$), in
contrast to the case of quantum turbulence. The same can be said about the
distribution of the loops in space of there lengths. In the theory of
quantum turbulence, the loop sizes are gathered on scales close to the
intervortex distance $1/\mathcal{L}.$

\subparagraph{The Lamb impulse}

Let's study now the Lamb impulse $\mathbf{P}$ and the effects associated
with it. The quantity $\left\langle \mathbf{Pv}_{ns}\right\rangle $ can be
evaluated from (\ref{Z1fin}), applying it to the velocity-dependent part of
PF $Z_{1Vns}$. In Fig.2 there are shown the quantity $\left\langle \mathbf{Pv%
}_{ns}\right\rangle /v_{ns}^{2}$ and energy $\left\langle E(\{\mathbf{s\})}%
\right\rangle /v_{ns}^{2}$ \ \ due to extra vortices induced by counterflow.
It is seen that they are of the same order as it should be. Formally, the
Lamb impulse $\mathbf{P}$ and the flow induced by it appear due to the
nonzero polarization of the extra vortex loops arising from counterflow.
\begin{figure}[tbp]
\includegraphics[width=7cm]{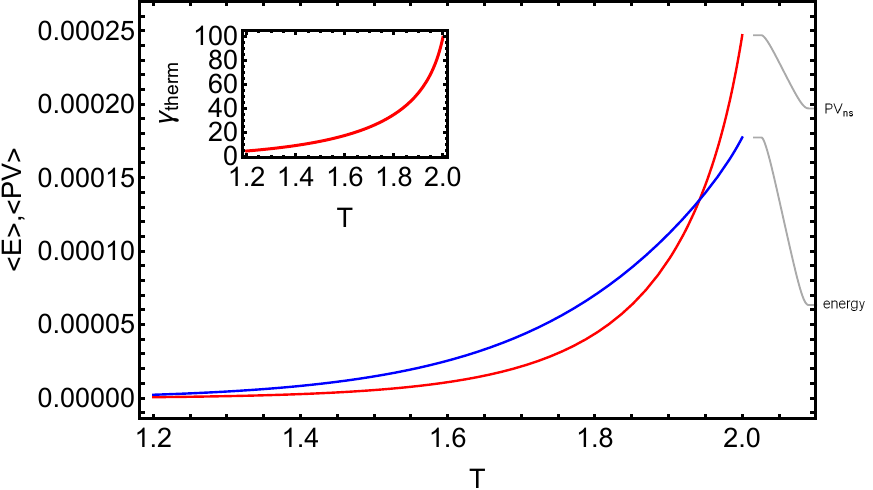}
\caption{(Color online) Quantities $\left\langle \mathbf{Pv}%
_{ns}\right\rangle /v_{ns}^{2}$ and energy $\left\langle E(\{\mathbf{s\})}%
\right\rangle /v_{ns}^{2}$. The inset shows the quantity $\mathcal{\protect\gamma }%
_{therm} $ \ as a function of temperature (see text)}
\label{PE}
\end{figure}

\subparagraph{Suppression of the superfluid density $\Delta \protect\rho %
_{s}(\mathbf{v}_{ns})$}

An analysis of the average $\left\langle \mathbf{Pv}_{ns}\right\rangle $
shows that the VT induces the superfluid flow directed against the external
superfluid $\mathbf{v}_{s}$. From macroscopic point of view, in particular
in hydrodynamic experiments, this additional superfluid mass current which
cancels the part of applied external velocity should display itself as a
suppression of the superfluid density $\Delta \rho _{s}(\mathbf{v}_{ns})$.
Depletion of the superfluid density $\Delta \rho _{s}(\mathbf{v}_{ns})$ in
counterflow was predicted in Landau's pioneering work \cite{Landau1941a} It
was calculated by Khalatnikov \cite{Khalatnikov2000} on the base of the
phonon and roton model. The results are that $\Delta \rho _{s}(\mathbf{v}%
_{ns})/\rho _{s}(0)$ are estimated as some numerical factor multiplied by $%
v_{ns}^{2}/c_{1}^{2}$ for phonons ($c_{1\text{ }}$is the first sound
velocity), and $v_{ns}^{2}/(k_{B}T/p_{0})^{2}$ for rotons ( $p_{0}$ is the
momentum of the roton minimum). In paper \cite{Ginzburg1979} the velocity
dependence of the superfluid density near the $\ \lambda $-point \ was
investigated. The result is that $\Delta \rho _{s}(\mathbf{v}_{ns})/\Delta
\rho _{s}(0)$ scales as $v_{ns}^{2}/(\kappa /a_{\lambda })^{2}$ ($a_{\lambda
}$ is the correlation length, singular neat $T_{\lambda }$ ).\newline
Let's study what the vortex model gives us. The quantity $\left\langle
\mathbf{Pv}_{ns}\right\rangle $ can be interpreted as $\Delta \rho _{s}(%
\mathbf{v}_{ns})v_{ns}^{2}$. Developing $\left\langle \mathbf{Pv}%
_{ns}\right\rangle $, extracted from (\ref{Z1fin}) in powers of velocity $%
v_{ns}$ we get the first nonvanishing term is $\propto v_{ns}^{2}$,
therefore there is a velocity independent contribution into $\Delta \rho
_{s} $. The numerical value of $\Delta \rho _{s}$can be extracted directly
from Fig.2. This contribution appears since \ excessive (due to counterflow)
vortices may be considered as quasiparticles (additional to rotons and
phonons), contributing into the superfluid density (and other thermodynamic
properties, such as heat capacity).\newline
The next term of the expansion (fourth order $v_{ns}^{2}$) corresponds to
the $v_{ns}^{2}$ dependence discussed above. One can be obtain that the
relative shift $\Delta \rho _{s}(\mathbf{v}_{ns})/\rho _{s}$ scales as $%
v_{ns}^{2}/(\kappa /l_{av})^{2}$, where $l_{av}$ is average size of loop,
close to correlation length $a_{\lambda }$. Literally, this result is close
to the one proposed in \cite{Ginzburg1979}, although with a small numerical
factor.

\subparagraph{Shift T$_{\protect\lambda }$}

There is a number of works, studying an impact of the counterflow on the
shift $T_{\lambda }$ (see, e.g. review \cite{Goodstein2003}). The physics of
this process is quite obvious. The kinetic energy of the counterflow changes
the free energy of the entire system, which naturally leads to a shift in
the transition point. This scheme had been used in various theoretical
models, such as the so called model F \cite{Hohenberg1977},\cite{Onuki1983},
the phenomenological theory of superfluidity of helium near the $\lambda $
point \cite{Ginzburg1976a}, a dynamic new look at the lambda transition \cite%
{Goodstein2003}. On the basis of these considerations as well as on
experimental data, cumulative formula was proposed for the relative shift $%
T_{\lambda }-T_{cr}$ ,%
\begin{equation}
(T_{\lambda }-T_{cr})/T_{\lambda }=(q/q_{0})^{1/2\nu }.  \label{shift other}
\end{equation}%
Here, the index $\nu =2/3$ is the crucial quantity in the theory of the
lambda transition , and $q_{0}$ spans from $1000$ W/cm$^{2}$ up to $10000$
W/cm$^{2}$.\newline
Let's consider the impact of the studying here vortex model on the shift of
the $\lambda $ point. The free energy $F$ of vortices is related to the one
loop PF as $F=-k_{B}T\ln Z=-k_{B}TZ_{1}$ (see (\ref{ZviaZ1})). The quantity $%
Z_{1}$ consists of velocity-free PF $Z_{1free}$ and the velocity dependent
part $Z_{1Vns}$. The former leads to free energy, which is positive and the
latter, associated with counterflow is negative. Accordingly, a phase
transition occurs when they balance each other. To move on we have to take
all the parameters entering Eq. (\ref{Z1fin}) in vicinity of critical
temperature $T_{cr}$ . We adopt the following relations for the superfluid
density and the correlation length near $T_{\lambda }$(see, e.g., \cite%
{Ginzburg1976a}, \cite{Goodstein2003})%
\begin{equation}
a_{\lambda }=4.75\ast 10^{-8}\ (T_{cr}-T)^{-2/3},\ \ \ \rho _{s\lambda
}=0.2\ (T_{cr}-T)^{2/3}.  \label{a rho lambda}
\end{equation}%
Using these data we get $Z_{1free}\propto (T_{cr}-T)^{2}$ which agrees with
the behaviour of free energy near $T_{\lambda }$ in the resting He II \cite%
{Ginzburg1976a}. Thus, at $T_{cr}$ the free energy associated with the
vortex system vanishes as it should be. The velocity dependent part of PF $%
Z_{1Vns}$ changes the whole free energy. To move further we rid of the
relative velocity $v_{ns}$, expressing it via the heat flux $q$ with the use
$v_{ns}=\rho q/\rho _{s}ST$. Substituting then the values $a_{\lambda },\ \
\rho _{s\lambda }$(\ref{a rho lambda}) into $Z_{1Vns}$, we get $%
Z_{1Vns}\propto q^{2}(T_{cr}-T)^{-2/3}$. Equating \ the obtained values $%
Z_{1free}$ and $Z_{1Vns}$, and taking into account that at zero counterflow $%
q=0$, $T_{cr}$ coincides with $T_{\lambda }$ we finally have (see Fig. (\ref%
{shift}))
\begin{equation}
q=2.83\ast 10^{3}(T_{\lambda }-T_{cr})^{4/3}.  \label{shift my}
\end{equation}%
This relation agrees with the cumulative formula (\ref{shift other}). As
both theoretical estimates and experimental data, although they give close
results with respect to the exponent ($4/3$), lead to a fairly large scatter
in the heat flux $q_{0}$. One of the possible reasons for this discrepancy
is related to the difficulty of interpreting experimental data, in which
both heat transfer and gravity have a significant effect.
\begin{figure}[tbp]
\includegraphics[width=7cm]{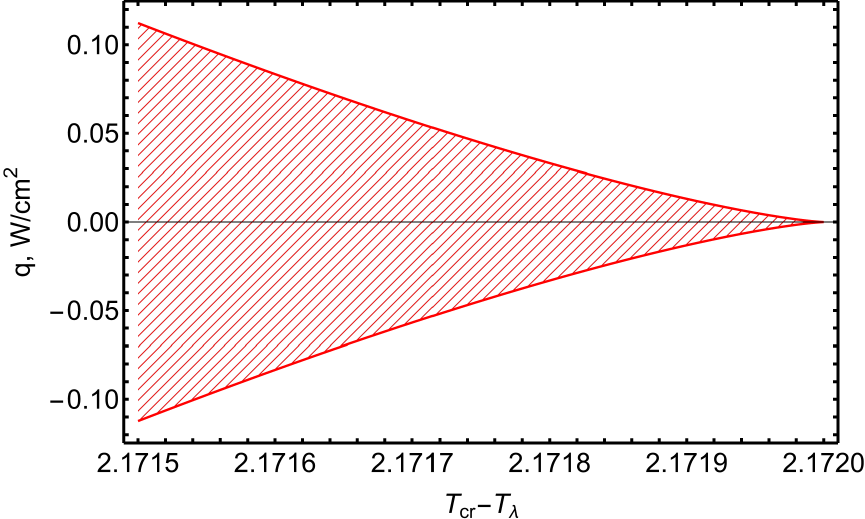}
\caption{(Color online) Shift T$_{\protect\lambda }$ as a function of
applied heat flux (see Eq. \protect\ref{shift my}). Superfluid phase is inside
hatched area.}
\label{shift}
\end{figure}

\section{Discussion and Conclusion}

The problem of thermodynamic quantized vortices in the counterflowing
superfluids, has been studied. Based on the calculated PF, some
properties of superfluids due to counterflow  were properly investigated. The
results are in good agreement with the data obtained on the basis
of other formalisms and from experiment.\\
The first important conclusion that follows from this is that this can be
considered a fairly strong argument in favor of the point of view that
the gas of topological excitations near the phase transition can indeed  play
the same role of quasiparticles as photons and rotons at low temperatures. We
refrain here from a tempting discussion of the question of the possible
connection between rotons and brownian vortex loops.\newline
One more conclusion, having important physical consequence,
is that there exists another population of vortex loops, which we call
hydrodynamic vortices. Their evolution is mainly due to the relative
velocity $\mathbf{v}_{ns}$, entering into Gibbs distribution (\ref{Gibbs}) -
(\ref{P}) as well as into partition function (\ref{Z1fin}). From the form of
the Hamiltonian $H\{\mathbf{s\}}$ it follows that this population is
concentrated near sizes where the energy $E\{\mathbf{s\}}$ competes with the
Lamb impulse $\mathbf{P}$ multiplied by $\mathbf{v}_{ns}$. Using the saddle
point approximation one can conclude that most favoured scale both size
loop size and for the mean radius of curvature is of the order $\kappa
/v_{ns}$. This result is close to what takes place in the theory of
superfluid turbulence - a very challenging topic, actively discussed at
present. Our result gives hope to understand the principal mechanisms
responsible for the formation the structure of the vortex tangle in
the stochastic regime. This question, as well as other
questions concerning the relationship of thermodynamic equilibrium vortices
with the quantum turbulence, is of great interest and will be investigated
in the future.\\

Study of Gibbs distribution and partition function had been conducted under
state contract with IT SB RAS (No. 17--117022850027--5), study of physical
applications was financially supported by the Russian Science Foundation
(Grant No. 23-22-00128).


%

\end{document}